\documentclass{article}
\usepackage{hyperref,amsfonts,amssymb, amsmath,mathrsfs}
\usepackage[english]{babel}

\newtheorem{prop}{Theorem}

\begin{document}


\title{Symbolic software for separation of variables in the Hamilton-Jacobi equation
for the L-systems.}
\author{Yu.A. Grigor{y}ev and
A.V. Tsiganov\\
\\
\it\small Department of Computational Physics, \\
\it\small St.Petersburg State University, St.Petersburg, Russia\\
\it\small e--mail: tsiganov@mph.phys.spbu.ru}
\date{}
\maketitle

\begin{abstract}
We discuss computer implementation of the known algorithm of finding
separation coordinates for the special class of orthogonal separable
systems called L-systems or Benenti systems.
\end{abstract}

\section{Introduction}
\setcounter{equation}{0}

Let $\mathcal Q$ be a$n$-dimensional Riemannian manifold with
generic local coordinates $q=(q^1,q^2,\ldots,q^n)$ and
positive-definite metric tensor $\mathbf G$.

On the cotangent bundle $T^*\mathcal Q$ of $\mathcal Q$ with
canonical coordinates $(p,q)$ we consider dynamical system with the
natural Hamilton function
\begin{equation} \label{nH-ben}
H=T(p,q)+V(q)=\sum_{i,j=1}^n \mathrm g^{ij}(q) p_ip_j+V(q).
\end{equation}
Here $\mathrm g^{ij}(q)$ are components of the metric tensor
$\mathbf G$ and $V(q)$ is the potential energy, a smooth function on
$\mathcal Q$ canonically lifted to a function on $T^*\mathcal Q$.

One of the most effective ways of solving the corresponding
equations of motion is by separation of variables in the
Hamilton-Jacobi equation
\begin{equation}\label{HJ-Eq}
H(p,q)=E.
\end{equation}
A coordinate system $Q=(Q^1,\ldots, Q^n)$ is called separable if the
Hamilton-Jacobi equation admits a complete solution of the form
\[
\mathcal S(Q,\alpha)=\sum_{i=1}^n \mathcal S_i(Q^i,\alpha), \qquad
\det\left[\frac{\partial^2\mathcal S}{\partial Q^i\partial
\alpha^j}\right]\neq 0\,.
\]
Here $\alpha=(\alpha^1,\ldots,\alpha^n)$ is a set of separation
constants. The corresponding Jacobi equations
\[
P_i\equiv\partial_i\mathcal S=\frac{\partial \mathcal
S_i(Q^i,\alpha)}{\partial Q^i}
\]
are called the separated equations. In a similar manner, we will
call a natural Hamiltonian or potential separable if such a
separable coordinate system exists.

For a given Hamilton function $H(p,q)$ the problem of finding
canonical transformation from initial variables $(p,q)$ to the
separation coordinates $(P,Q)$ is very non-trivial. The problem was
originally formulated by Jacobi when he invented elliptic
coordinates and successfully applied them to solve several important
mechanical problems, such as the problem of geodesic motion on an
ellipsoid and the Euler problem of planar motion in a force field of
two attracting centers and the problem of the motion of three
particles, which interact due to forces depending on their relative
distances \cite{jac36}.

Up to now  finding  the separation coordinates for a given
integrable system remains  rather the magic art than a constructive
theory. However, for the special class of the natural Hamiltonians
we have a complete and algorithmic solution of this problem
\cite{eis34,ka80, ben93,rw04}. The algorithm is straightforward
enough to be implemented on a computer and thus turned into a
practical tool.

In this note we present an implementation  of this algorithm by
means of the computer algebra system Maple 9.5. The corresponding
Maple file with the code may be found in
\begin{verbatim}
http://www.maplesoft.com/applications/app_center_view.aspx?AID=1686
\end{verbatim}

\section{Algorithm of the point separation of variables}
According to \cite{eis34,ben93} we restrict ourselves to the search
of the point canonical transformations $Q=f(q)$ and $P=g(q,p)$ only.

In this case from St\"ackel (1893), Levi-Civita (1904), Eisenhart
(1934), Kalnins \& Miller (1980) and Benenti (1993)
 we have
\begin{prop}
The Hamilton-Jacobi equation (\ref{HJ-Eq}) is separable in
orthogonal coordinates if and only if there exists symmetric Killing
2-tensor $\mathbf K$ with simple eigenvalues and normal eigenvectors
such that
\begin{equation}\label{char-eq}
 d(\mathbf K dV ) = 0,
\end{equation}
 where $d$ denotes
the exterior derivative.
\end{prop}
Such Killing tensor $\mathbf K$ is called characteristic, its
existence is completely defined by the kinetic energy $T$. It means
that the separation of the geodesic equation is a necessary
condition for the separation of equation $T+V=E$. The equation
$d(\mathbf K dV ) = 0$ is an integrability condition for the
existence of the potential $V(q)$ that may be added to $T$.

In 1992 Benenti  has shown a simple  recurrence procedure to
construct a special family of Killing tensors $\mathbf K$ obeying
the assumptions of this theorem. He considered a special class of
Riemannian manifolds $\mathcal Q$ endowed with the L-tensor,  whose
functionally independent eigenvalues are identified with the
separation variables.

Following Benenti let us  call L-tensor a  conformal Killing tensor
$\mathbf L$ with vanishing torsion and  pointwise simple eigenvalues
$Q_i$.  Under these conditions the tensors
\begin{equation}\label{sigm-magri}
 \mathbf
K_m=\sum_{k=0}^{m}\sigma_{m-k}\mathbf L^k,\qquad\mbox{\rm or}\qquad
\mathbf K_m=\sigma_m\mathbf G-\mathbf K_{m-1}\mathbf L\,,\qquad
m=0,\ldots,n-1,
\end{equation}
where $\mathbf L$ is (2,0) tensor field, are the Killing tensors
with simple eigenvalues and normal eigenvectors.  Here functions
$\sigma_m$ are the elementary symmetric polynomials of degree $m$ on
the eigenvalues of $\mathbf L$, such that $\det (\lambda\mathrm
I-\mathbf L)=\displaystyle{\sum_{m=0}^n \sigma_m\lambda^{n-m}}$.

It can be shown \cite{ben93} that equation $d(\mathbf K dV ) = 0$
implies $d(\mathbf K_m dV ) = 0$ for all the commuting Killing
tensors. This allows to define the new potentials $V_m$ according to
\begin{equation}\label{pot-Vm}
dV_m=\mathbf K_mdV,\qquad m=1,\ldots,n-1
\end{equation} and to introduce integrals of
motion
\begin{equation}\label{int-Hm}
H_m=\sum_{i,j=1}^n \mathbf K^{ij}_{m}(q) p_ip_j+V_m(q),\qquad
m=1,\ldots,n-1.
\end{equation}
These functions form a family of commuting integrals of the motion
for the Hamiltonian $H$ (\ref{nH-ben}).

According to \cite{imm00} integrals of motion  $H_m$ (\ref{int-Hm})
are solutions of the recursion relations
\begin{equation}\label{recc-len-pedr}
dH_{m+1}=\mathbf N^*dH_m+\sigma_{m+1}dH,\qquad m=1,\ldots,n-1,\qquad
H_{n}\equiv0\,.
\end{equation}
Here $\mathbf N$ is the recursion operator, which is the complete
lifting of (1,1) tensor field $\mathbf L$ to $T^*\mathcal Q$
\[
\mathbf N\,\dfrac{\partial }{\partial  q^k}=\sum_{i=1}^n L_k^i
\dfrac{\partial }{\partial  q^i}+\sum_{ij} p_j\left(\dfrac{\partial
L^j_i}{\partial q^k}-\dfrac{\partial  L^j_k}{\partial
q^i}\right)\dfrac{\partial }{\partial  p_i},\qquad \mathbf
N\,\dfrac{\partial }{\partial  p_k}=\sum_{i=1}^n L^k_i
\dfrac{\partial }{\partial p_i}.
\]
Recall, since a metric tensor is present, the boldface object
$\mathbf K$ or $\mathbf L$ can be represented in components as a
tensor of type (2, 0), (1, 1) and (0, 2), respectively.

So, in order to get separation variables and integrals of motion for
a given integrable system we have to find tensor $\mathbf L$ only.
As noticed in \cite{cr03}, tensor $\mathbf L$ is an L-tensor with
respect to the usual Riemannian metric $\mathbf G$ iff
\begin{equation}
d(\mathcal L_{X_T}\,\theta-T d\sigma_1)=0,\label{maple1}
\end{equation}
where $\mathcal L$ is the Lie derivative along the geodesic vector
field $X_T$, $\sigma_1=\mbox{\rm tr}\,\mathbf L$ is first symmetric
polynomial and
\begin{equation}\label{theta}
 \theta=\sum_{i,j=1}^n
L^i_j\,p_idq^j
\end{equation}
is the L-deformation of the standard
Liouville 1-form $\theta_0=\sum p_j dq^j$ for any set of fibered
coordinates $(p,q)$.

The equation $d(\mathbf K dV ) = 0$ with the Killing tensor $\mathbf
K_1$ (\ref{sigm-magri}) may be rewritten in the similar form
\begin{equation}
d(\mathcal L_{X_V}\,\theta-Vd\sigma_1)=0\,,\label{maple2}
\end{equation}
see for instance \cite{bfp03}.

Below we will use the following well-known expression for the Lie
derivative $\mathcal L$ along the vector field $X$
\[
\mathcal L_X=\mathtt{i}_X\,d+d\,\mathtt{i}_X\,.
\]
Here $\mathtt{i}_X$ is a hook operator and $d$ is an exterior
derivative. It allows us to decrease the amount of intermediate
calculations  as long as $d^2=0$ and we have
\[d\mathcal L_X=d\,\mathtt{i}_X\,d+d^2\,\mathtt{i}_X=d\,\mathtt{i}_X\,d.\]
In these notations equations (\ref{maple1}) and (\ref{maple2}) read
as
\begin{eqnarray}
d(\mathtt{i}_{X_T}d\,\theta-Td\sigma_1)&=&0,\label{maple3}\\
d(\mathtt{i}_{X_V}d\,\theta-Vd\sigma_1)&=&0\,.\label{maple4}
\end{eqnarray}

We will call L-system or Benenti system any separable orthogonal
system whose Killing tensor $\mathbf K$ in (\ref{char-eq}) is
generated by an L-tensor according to (\ref{sigm-magri}). To
construct the separation coordinates $Q$ for the  L-system we have
to solve equations (\ref{maple3}) and (\ref{maple4}) with respect to
functions $L^i_j(q)$ and to find the eigenvalues of the tensor
$\mathbf L$.

In the next section we present a computer program for search of the
L-tensors and the associated separation coordinates for L-systems.

\vskip0.3truecm\par\noindent \textbf{Remark:} The 1-form $\theta$
may be used to construct the second Poisson structure on
$T^*\mathcal Q$ \cite{imm00,bfp03}. In this case the recursion
operator $N$ will be a complete lifting of the L-tensor $\mathbf L$
from the configuration space to the whole phase space. Therefore,
the coordinate separation of variables can be  treated as a
particular case of the bi-hamiltonian theory of separation of
variables.

\section{The program for the search of separation variables}
In this section we present an implementation  of the discussed
 algorithm made in the symbolic computational system Maple
v.9.5.

We will use the standard Maple package \verb" liesymm", which was
designed for construction of the differential forms corresponding to
partial differential equations. In fact we are doing the inverse
procedure, i.e. starting with the differential forms
(\ref{maple3}-\ref{maple4}) we have to get a system of partial
differential equations.

Let us start with the following command
\begin{verbatim}> with(liesymm):\end{verbatim}
which makes the short form names of the functions of a Maple package
available at the interactive level.

At the first step we need to determine  dimension of a given
configuration space $\mathcal Q$ and to suppose that the phase space
$T^*\mathcal Q$ is equipped with some canonical coordinates
$q=(q^1,\ldots,q^n)$ and $p=(p_1,\ldots,p_n)$:
\begin{verbatim}
> n := 2;
\end{verbatim}
\[
n:= 2
\]
\begin{verbatim}> q:=seq(q||i,i=1..n): p:=seq(p||i,i=1..n): var:=q,p:
> setup(var);
\end{verbatim}
\[
[ q1, \, q2, \, p1, \, p2]
\]
Now we will  look for a tensor field $\mathbf L$ with vanishing
torsion and functionally independent eigenvalues. For a given
integrable system tensor L has to satisfy equations (\ref{maple3})
and (\ref{maple4}) which include first symmetric polynomial
$\sigma_1$ on the eigenvalues of $\mathbf L$
\begin{verbatim}> sigma:=add(L[i,i](q),i=1..n);\end{verbatim}
\[
\sigma := {L_{1, \,1}}( q1, \, q2) + {L_{2, \,2}}( q1, \, q2)
\]
and L-deformation $\theta$ (\ref{theta}) of the standard  Liouville
form
\begin{verbatim}
> theta:=add(add(L[i,j](q)*p||i*d(q||j),i=1..n),j=1..n);
\end{verbatim}
\begin{eqnarray}
\theta := {L_{1, \,1}}( q1, \, q2)\, p1 \,\mathrm{d}( q1) + {L_{2,
\,1}}( q1, \, { q2})\, p2\,\mathrm{d}( q1) \nonumber\\+ {L_{1,
\,2}}(  q1, \, q2)\, p1\,\mathrm{d}( q2) + {L_{2, \,2}}( q1, \,
q2)\, p2\, \mathrm{d}( q2)\nonumber
\end{eqnarray}
Here  components of L-tensor $L^i_j(q)$ are the unknown functions on
the configuration space $\mathcal Q$. This L-tensor gives rise to a
torsionless tensor field $\mathbf N$ of type (1,1) on T*Q, which
acts on the 1-form of fibered coordinates $(p,q)$ as
\begin{verbatim}
> Nq:={seq(d(q[k])=add(L[k,j](q)*d(q[j]) ,j=1..n),k=1..n)}:
> Np:={seq(d(p[k])=add(L[j,k](q)*d(p[j])
> - p||j*add((diff(L[j,i](q),q[k]) -diff(L[j,k](q),q[i]))*d(q||i),i=1..n),
>      j=1..n),k=1..n)}:
\end{verbatim}

At the second step we introduce canonical Poisson tensor $\mathscr
P$
\begin{verbatim}
> ed:=array(identity, 1..n,1..n): u:=array(sparse,1..n,1..n):
> P:=linalg[stackmatrix](linalg[augment](u,ed),
                         linalg[augment](-ed,u));
\end{verbatim}
\[
P :=  \left[ {\begin{array}{rrrr}
0 & 0 & 1 & 0 \\
0 & 0 & 0 & 1 \\
-1 & 0 & 0 & 0 \\
0 & -1 & 0 & 0
\end{array}}
 \right]
\]
and canonical Poisson brackets
\begin{verbatim}
> PB:=proc(f,g) options operator, arrow:
> add( diff(f,p||i)*diff(g,q||i)- diff(f,q||i)*diff(g,p||i), i=1..n)
> end:
\end{verbatim}
It allows us to calculate the vector fields $X_T=\mathscr P dT(p,q)$
and $X_V=\mathscr P dV(q)$. They are equal to
\begin{verbatim}
> dT:=array(1..2*n):
> for i from 1 to 2*n do dT[i]:=diff(T(var),var[i]): end do:
> X_T:=evalm(P&*dT);
\end{verbatim}
\begin{eqnarray}
\mathit{X T} :=  &\Bigl[& {\frac {
\partial }{\partial \mathit{p1}}}\,\mathrm{T}(\mathit{q1}, \,
\mathit{q2}, \,\mathit{p1}, \,\mathit{p2}), \,{\frac {\partial }{
\partial \mathit{p2}}}\,\mathrm{T}(\mathit{q1}, \,\mathit{q2}, \,
\mathit{p1}, \,\mathit{p2}), \nonumber\\
 &-&({\frac {\partial }{\partial \mathit{q1}}}\,\mathrm{T}(
\mathit{q1}, \,\mathit{q2}, \,\mathit{p1}, \,\mathit{p2})), \, -
({\frac {\partial }{\partial \mathit{q2}}}\,\mathrm{T}(\mathit{q1 },
\,\mathit{q2}, \,\mathit{p1}, \,\mathit{p2})) \Bigr]\nonumber
\end{eqnarray}
and
\begin{verbatim}
> dV:=array(1..2*n):
> for i from 1 to 2*n do dV[i]:=diff(V(q),var[i]): end do:
> X_V:=evalm(P&*dV);
\end{verbatim}
\[
 {X V} :=  \left[  \! 0, \,0, \, - ({\frac {\partial }{
\partial  q1}}\,\mathrm{V}( q1, \, q2)),
\, - ({\frac {\partial }{\partial  q2}}\,\mathrm{V}(  q1, \, q2)) \!
\right]
\]
The equations with the vector field $X_V$ will be simpler due to its
zero components than ones with $X_T$ and, therefore, we
 start with the second equation (\ref{maple4}) in
order to explain some features of the Maple procedures. Namely,
substituting the vector field $X_V$ and 1-form $\theta$ into the
 $\mathtt{i}_{X_V}\,d\theta$, one gets
\begin{verbatim}
> idT:=wcollect(hook(d(theta),X_V)):
\end{verbatim}
\begin{eqnarray}
  idT := \Bigl(&-& {L_{2, \,1}}( q1, \, q2)\,(
{\frac {\partial }{\partial  q2}}\,\mathrm{V}( q1 , \, q2))( q1, \,
q2, \, p1, \,
 p2)\nonumber\\
&-& {L_{1, \,1}}( q1, \, q2)\,({\frac {
\partial }{\partial  q1}}\,\mathrm{V}( q1, \,
 q2))( q1, \, q2, \, p1, \,
 p2)\Bigr)\mathrm{d}( q1)\nonumber \\+\Bigl(
&-& {L_{2, \,2}}( q1, \, q2)\,({\frac {\partial }{
\partial  q2}}\,\mathrm{V}( q1, \, q2))(
 q1, \, q2, \, p1, \, p2) \nonumber\\
&-& {L_{1, \,2}}( q1, \, q2)\,({\frac {
\partial }{\partial  q1}}\,\mathrm{V}( q1, \,
 q2))( q1, \, q2, \, p1, \,
 p2)\Bigr)\mathrm{d}( q2)\nonumber
\end{eqnarray}
It is easy to see that this expression contains derivatives of the
potential $V(q)$, which formally depend on all the variables
$(q,p)$. In fact these derivatives depend on the variables $q$ only.
It is  an unpleasant feature of the Maple procedure
\verb"hook" from the package
\verb"liesymm".

In order to get the correct expression for
$\mathtt{i}_{X_V}\,d\theta$ we have to use the special substitution
\begin{verbatim}
> Trans:={seq(diff(V(q),q||i)(var)=diff(V(q),q||i),i=1..n)}:
> idT:=subs(Trans,idT);
\end{verbatim}
\begin{eqnarray}
  idT := \Bigl(&-& {L_{2, \,1}}( q1, \, q2)\,
{\frac {\partial }{\partial  q2}}\,\mathrm{V}( q1 , \, q2)-{L_{1,
\,1}}( q1, \, q2)\,{\frac {
\partial }{\partial  q1}}\,\mathrm{V}( q1, \,
 q2)\Bigr)\mathrm{d}( q1)\nonumber \\+\Bigl(
&-& {L_{2, \,2}}( q1, \, q2)\,{\frac {\partial }{
\partial  q2}}\,\mathrm{V}( q1, \, q2)- {L_{1, \,2}}( q1, \, q2)\,{\frac {
\partial }{\partial  q1}}\,\mathrm{V}( q1, \,
 q2)\Bigr)\mathrm{d}( q2)\nonumber
\end{eqnarray}
The equation (\ref{maple4}) are satisfied identically and
coefficients for independent 2-forms in (\ref{maple4}) vanish. It
gives rise to the following system of equations
\begin{verbatim}
> SysEq:=annul(d(idT-V(q)*d(sigma)), [var] ): nops(SysEq);
\end{verbatim}
At $n=2$ equation (\ref{maple4}) generates  only one partial
differential equation, while at $n=3$  the system
\verb"SysEq" consists of three equations and so on.

We keep these equations  in a special list
\begin{verbatim}
> ListEq:=NULL:
> for i from 1 to nops(SysEq) do
>     ListEq:=ListEq,lhs(SysEq[i]);
> end do:
\end{verbatim}
Now we pass on the remaining equation (\ref{maple3}). As above we
have to simplify this equation through the additional substitution
for derivatives after the application of the
\verb"hook" operator
\begin{verbatim}
> Eq:=wcollect(d(hook(d(theta),X_T)-T(var)*d(sigma))):
> Trans:={seq( diff(T(var),var[i])(var)=diff(T(var),var[i]),i=1..2*n)}:
\end{verbatim}
Expanding equation (\ref{maple3}) in the basis of 2-forms one gets
one more system of equations
\begin{verbatim}
> SysEq:=annul(subs(Trans,Eq),[var]): nops(SysEq);
\end{verbatim}
At $n=2,3$ equation (\ref{maple4}) gives rise to six and fifteen
equations respectively.

By adding these equations to earlier prepared list of equations
\verb"ListEq" one gets a complete set of algebraic and partial
differential equations on the components of L-tensor.
\begin{verbatim}
> for i from 1 to nops(SysEq) do ListEq:=ListEq,lhs(SysEq[i]); end do:
> NumEq:=nops([ListEq]);
\end{verbatim}
\[
 {NumEq} := 7
\]
At $n=2,3$ one gets seven  and eighteen  equations on four and nine
functions $L^i_j(q)$ respectively.

At the third step we have to define the Hamilton function. As an
example we consider the H\'{e}non-Heiles integrable model with the
 Hamiltonian
\begin{equation}\label{HH-ham}
H=p_1^2+p_2^2+\frac12(q_1^2+q_2^2)+q_1^2q_2+2q_2^3\,.
\end{equation}
In this case kinetic energy $T$ and potential $V(q)$ are given by
\begin{verbatim}
> Tn:=add(p||i^2,i=1..n);
> Vn:=1/2*(q1^2+q2^2)+q1^2*q2+2*q2^3;
\end{verbatim}
Substituting these parts of the Hamilton function into the
 prepared above list of
equations \verb"ListEq" one gets a set of polynomial equations of
the second degree in the momenta $p$, which must be identically
satisfied for all admissible values of  variables $p$.  It means
that coefficients for the second, first and zeroth power of $p_i$
vanish.  All these coefficients form  new system of algebraic and
partial differential equations on the functions $L^i_j(q)$.
\begin{verbatim}
> System:= NULL:
> for i from 1 to NumEq do
>     u:=expand(dvalue(subs({ T(var)=Tn, V(z)=Vn },ListEq[i]))):
>     System:=System,coeffs(u,{p}):
> end do:
> NumSys:=nops({System});
\end{verbatim}
\[NumSys:=13\]
At $n=2,3,4$ the resulting system of equation  consists of 13, 51
and 136 equations. Even though some of these equations may be
dependent, it is convenient to use them all simultaneously. Of
course, for a generic potential this system is heavily
overdetermined and has only the trivial solution, which means that
the Hamilton function is non-separable.

In the case of polynomial potentials we can easy solve this system
by means of the standard Maple procedure
\verb"pdsolve"
\begin{verbatim}
> Ans:=pdsolve( {System}, {seq(seq(L[i,j](z), i=1..n),j=1..n)} );
\end{verbatim}
which allows us to get the following answer
\begin{eqnarray}
Ans:= \{{L_{1, \,1}}(\mathit{q1}, \,\mathit{q2})= {\displaystyle
\frac {3\,C_1}{4}}  + C_2, \,{ L_{1, \,2}}(\mathit{q1},
\,\mathit{q2})={\displaystyle \frac {
C_1\,\mathit{q1}}{2}} ,  \nonumber\\
{L_{2, \,2}}(\mathit{q1}, \,\mathit{q2})=C_1\,\mathit{ q2} + C_2,
\,{L_{2, \,1}}(\mathit{q1}, \,\mathit{q2})= {\displaystyle \frac
{C_1\,\mathit{q1}}{2}} \}  \nonumber
\end{eqnarray}
 depending on two arbitrary constants
$C_{1,2}$.

Substituting this answer into the L-tensor one gets
\begin{verbatim}
> Trans:={seq(seq(L[i,j](q)=L[i,j],i=1..n),j=1..n)}:
> Ans:=simplify(map2(subs,Trans,Ans)):
> L:=array(1..n,1..n):
> L:=map2(subs,Ans,evalm(L));
\end{verbatim}
\[\mathit{L} :=  \left[ {\begin{array}{cc} {\displaystyle  \frac
{3}{4}}\,C_1  + C_2 & {\displaystyle \frac {\mathit{C_1q1}}{2}}
\\ [2ex] {\displaystyle \frac{\mathit{C_1q1}}{2}} & \mathit{
 C1}\,\mathit{q2} + C_2
\end{array}}
 \right]
\]
The eigenvalues of this L-tensor are the separation coordinates
\begin{verbatim}
> Q:=linalg[eigenvalues](Ln);
\end{verbatim}
\begin{eqnarray}
\mathit{Q}:=&&\frac {3C_1}{8}  + C_2 + \frac {C_1\,\mathit{q2}}{2} +
\frac{\sqrt{ 9\,C_1^{2} - 24\, C_1^{2}\,\mathit{q2} +
16\,C_1^{2}\,\mathit{ q2}^{2} +
16\,C_1^{2}\,\mathit{q1}^{2}}}{8} ,\nonumber\\
&&\frac {3\,C_1}{8}  + C_2 + \frac {C_1\,\mathit{q2}}{2} - \frac
{\sqrt{9\,C_1^{2} - 24\, C_1^{2}\,\mathit{q2} +
16\,C_1^{2}\,\mathit{ q2}^{2} + 16\,C_1^{2}\,\mathit{q1}^{2}}}{8}
.\nonumber
\end{eqnarray}
Thus one gets translated parabolic coordinates, which are the
separation coordinates for the H\'{e}non-Heiles  model \cite{rw04}.
It takes less then a minute of the  computer time.

More explicitly, we obtain a family of equivalent separable
coordinate systems labeled by $C_{1,2}$. Recall, that two separable
systems are called equivalent if the corresponding separated
solutions of the Hamilton-Jacobi equation generate the same
Lagrangian foliation of $T^*\mathcal Q$.

Now let us consider construction of the corresponding  integrals of
motion in the framework of the Riemannian geometry. Following
Benenti we have to construct polynomials $\sigma_m$
\begin{verbatim}
> for i from 0 to n do
>   sigma[i]:=coeff(linalg[det](lambda*ed-L),lambda,n-i):
> end do:
\end{verbatim}
and basis of Killing tensors $\mathbf K_m$ (\ref{sigm-magri})
\begin{verbatim}
> for m from 1 to n-1 do
>   K||m:=evalm(add( sigma[m-k]*L^k,k=0..m)):
> end do:
\end{verbatim}
Remind, that in our case $\mathrm g^{ij}=1$ and, therefore,
$L^{ij}=L^i_j$.

In addition to $\mathbf K_m$ equations (\ref{pot-Vm}) consist of
exterior derivatives on the unknown potentials $dV_m$, which  are
defined by
\begin{verbatim}
> for m from 0 to n-1 do
>  dV||m:=array(1..n): for i from 1 to n do
>       dV||m[i]:=diff(V||m(q),q[i]):
>     end do:
> end do:
> dV0:=map2(subs,V0(q)=Vn,dV0):
\end{verbatim}
Now we can solve equations (\ref{pot-Vm}) and determine integrals of
motion $H_m$ (\ref{int-Hm})
\begin{verbatim}
for m from 1 to n-1 do
>   eqV:=evalm(dV||m-K||m&*dV0):
>   Ans:=pdsolve({seq(eqV[i], i=1..n)},V||m(q)):
>   H[m]:=add(add( K||m[i,j]*p||i*p||j ,i=1..n),j=1..n)+subs(Ans,V||m(q)):
> end do:
\end{verbatim}
We can show the Hamilton function
\begin{verbatim}
> H[0]:=Tn+Vn;
\end{verbatim}
\[
{H_{0}} := \mathit{p1}^{2} + \mathit{p2}^{2} + {\displaystyle \frac
{\mathit{q1}^{2}}{2}}  + {\displaystyle \frac {\mathit{q2} ^{2}}{2}}
+ \mathit{q1}^{2}\,\mathit{q2} + 2\,\mathit{q2}^{3}
\]
and  prove that our integrals of motion $H_m$  commute with the
Hamiltonian $H_0$
\begin{verbatim}
> for m from 1 to n-1 do
>   ZERO:=simplify(PB(H[0],H[m]));
> end do;
\end{verbatim}
\[
\mathit{ZERO} := 0
\]

\vskip0.3truecm\par\noindent \textbf{Remark:}  At $n\leq 10$ and for
the polynomial potentials  we can solve  overdetermined system of
equation \verb"System"  with the standard procedure
\verb"pdsolve" on  a standard personal computer
in the reasonable time. At $n>10$ we have to use special computers
or special symbolic software for solving heavily overdetermined
systems of equations.

\section{Examples}
In this section we present some examples which illustrate various
aspects of the proposed code.

\subsection{The anharmonic oscillator}
Configuration space $\mathcal Q$ is the $n$-dimensional Euclidean
space $\mathbb R^n$ with cartesian coordinates $q$ such that metric
takes the form $ds^2=\sum\mathrm{g}_{ij}dq_idq_j=\sum d{q_i}^2$. Let
us consider the anharmonic oscillator with the
 Hamiltonian
\[
H=\sum_{i=1}^n p_i^2+\sum_{i=1}^n a_iq_i^2+\left(\,\sum_{i=1}^n
q_i^2\,\right)^2.
\]
We have to substitute this Hamiltonian instead of (\ref{HH-ham})
using commands
\begin{verbatim}
> Tn:=add(p||i^2,i=1..n);
> Vn:=add(a||i*q||i^2,i=1..n)+(add(q||i^2,i=1..n))^2;
\end{verbatim}
In is easy to calculate  solution of the complete system of
equations \verb"System" obtained from equations (\ref{maple3}) and
(\ref{maple4})
\[
{L}^i_j(q)=(C_1+a_i -a_n)\delta_{ij} + C_2\,q_iq_j
\]
Here the solution found by Maple is presented without any further
simplification.

At $C_1=a_n$ and $C_2=1$ the eigenvalues $Q_i$ of the corresponding
tensor $\mathbf L$ are defined  by the equation
\begin{equation}\label{ell-coord}
{\displaystyle
\frac{\det(L-\lambda)}{\prod_{i=1}^n(\lambda-a_i)}}\equiv-1+\sum_{i=1}^
n\frac{{q_i}^2}{\lambda-a_i}=\prod_{i=1}^n\frac{\lambda-Q_i}{\lambda-a_i}\,.
\end{equation}
This is the well-known relation which defines standard elliptic
coordinates in $\mathbb R^n$.

These coordinates were introduced by Jacobi in a note in Crelle's
Journal \cite{jac37}. A thorough discussion of its general
properties as well as of its use for separation of variables in the
Hamilton-Jacobi equation can be found in his lecture notes
\cite{jac36}.

Now let us consider construction of integrals of motion in the
involution in framework of the bi-Hamiltonian geometry. At first we
determine action of the recursion operator $\mathbf N$ on the
1-forms $dp$ and $dq$ using given tensor $\mathbf L$
\begin{verbatim}
> Nqn:=wcollect(simplify(map2(subs,Ans,Nq))):
> Npn:=wcollect(simplify(map2(subs,Ans,Np))):
\end{verbatim}
Then we calculate symmetric polynomials $\sigma_m$ on eigenvalues of
$\mathbf L$
\begin{verbatim}
> for m from 0 to n do
>  sigma[m]:=coeff(linalg[det](Ln-lambda*ed),lambda,n-m):
> end do:
\end{verbatim}
At the final step we construct and solve recursion relations
(\ref{recc-len-pedr})
\begin{verbatim}
> unassign(H): H[n+1]:=0: H[0]:=Tn+Vn;
> for m from n-1 by -1 to 1 do
>   Eq:=wcollect(dvalue( d(H[m+1])-
>       subs(Nqn union Npn,d(H[m](var)))-sigma[m+1]*d(H[0]))):
>   Sys:=annul(Eq,[var]):
>   Ans1:=pdsolve(Sys,H[m](var));
>   H[m]:=subs(Ans1,H[m](var)):
> end do:
\end{verbatim}
In addition we can check the Poisson brackets relations between
integrals of motion
\begin{verbatim}
> for m from 1 to n-1 do
>   ZERO:=simplify(PB(H[0],H[m]));
> end do;
\end{verbatim}

\subsection{The Euler two centers problem}
Let us consider the Euler problem of planar motion in a force field
of two attracting centers with the Hamiltonian
\[
H=p_1^2+p_2^2-\left(\frac{a_1}{\sqrt{(q_1-c)^2+q_2^2}}
+\frac{a_2}{\sqrt{(q_1+c)^2+q_2^2}}\right)\,.
\]
In contrast with the previous examples potential $V(q)$ is an
algebraic function instead of a polynomial one.

Solution of the overdetermined system of equations obtained from
(\ref{maple3}) and (\ref{maple4}) is equal to
\[
L=\left[\begin{array}{cc}(q_1^2-c^2)C_2+C_1& q_1q_2C_2\\
q_1q_2C_2&q_2^2C_2+C_1\end{array}\right]
\]
At $C_1=1/2\,c^2$ and $C_2=1$ the eigenvalues of the matrix $L$ are
the standard elliptic coordinates
\[
1-\frac{q_1^2}{\lambda+c^2/2}-\frac{q_2^2}{\lambda-c^2/2}=
\frac{(\lambda-Q_1)(\lambda-Q_2)}{(\lambda-c^2/2)(\lambda+c^2/2)}
\]

\subsection{The Toda lattice}
As above, configuration space $\mathcal Q$ is the $n$-dimensional
Euclidean space $\mathbb R^n$ with cartesian coordinates $q$. The
Hamiltonian of the periodic Toda lattice is given by
\begin{equation}
\label{hamTod} H=\frac12\sum_{i=1}^n
{p_i}^2+\sum_{i=1}^n\exp(q_i-q_{i+1})\ ,\qquad q_{n+1}\equiv q_1\,.
\end{equation}
At $n=2$ there exists the unique solution of the complete system of
equations \verb"System"
\[
L=\left[\begin{array}{cc}C_1 & C_2\\
C_2& C_1\end{array}\right]
\]
Of course, eigenvalues of this matrix are not the separation
variables. However, in this case the number of free parameters is
equal to the dimension of the Riemannian manifold $\mathcal Q$. It
allows us to construct the separation coordinates too (see
\cite{rw04}). Namely, let us diagonalize matrix $L$
\[ L=V^{-1}\mathrm{diag}(C_1+C_2,C_1-C_1)V,\qquad V=\left(%
\begin{array}{cc}
  1 & 1 \\
  -1 & 1
\end{array}%
\right)\,.
\]
It is easy to see that the separation coordinates $Q$ are the
following cartesian coordinates
\[
Q=Vq,\qquad \Rightarrow\qquad Q_1=q_1+q_2, \quad Q_2=q_1+q_2\,.
\]
At $n=3$ solution of the equations \verb"System" is the trivial o
\[
\mathit{L}=  \left[ {\begin{array}{ccc}
C_1 & C_2 & C_2 \\
C_2 & C_1 & C_2 \\ C_2 & C_2 & C_1
\end{array}}
 \right]
\]
The number of free parameters is less than the dimension of the
Riemannian manifold. It means that this Hamiltonian does not
separable  through  point transformations. However, we have to
underline that this Hamiltonian admits separation of variables in a
wider class of canonical transformations of the whole phase space.

\subsection{The Neumann system}
The unit sphere $\mathbb S^n$ is the Riemannian subspace of $\mathbb
R^{n+1}$ whose points have cartesian coordinates
$x=(x_1,\ldots,x_{n+1})$ satisfying $|x|=\sqrt{(x,x)}=1$.

The Neumann system is a well-known and very much studied mechanical
system, which describes  particle moving on the sphere under the
influence of a quadratic potential $V=\frac12\sum a_ix_i^2$. Due to
the form of the constraint and of the potential, it is quite natural
to perform all the computations in the local coordinates
$q_i=x_i^2$, $i=1,\ldots,n$. In these coordinates the Hamiltonian of
the Neumann system  is given by $H=T+V$ where
\[
T=2\sum_{i=1}^n q_i(1-q_i)p_i^2-4\sum_{i<j}^n q_iq_jp_ip_j,\qquad
V=\frac12\sum_{i=1}^n(a_i-a_{n+1})q_i\,.
\]
Here $p_i$ are  momenta conjugated to $q_i$ and $a_1,\ldots,a_{n+1}$
are arbitrary parameters \cite{bfp03}. We can insert this
Hamiltonian in the Maple code with the commands
\begin{verbatim}
> Tn:=2*add(q||i*(1-q||i)*p||i^2,i=1..n)
>    -4*add(add(q||i*q||j*p||i*p||j,j=i+1..n),i=1..n);
> Vn:=1/2*add((a||i-a||(N+1))*q||i,i=1..n);
\end{verbatim}
At $n=2$ the unique solution of the complete system of equations
\verb"System" looks like
\[
{L}= \left[\begin{array}{cc} C_2-{\displaystyle
\frac{{q_1}({a_3}-{a_1})+{a_1}-{a_2}}{{a_2}-{a_3}}} \,C_1 &
{q_1\,}C_1   \\[2ex]
  {\displaystyle \frac {{q_2}\,
({a_1} - {a_3})}{{a_2} - {a_3}}}\,C_1 & {q_2}\,C_1 + C_2
\end{array} \right]
\]
Recall, that $L^i_j(q)$ are components of the  tensor $\mathbf L$,
which is symmetric with respect to the metric $\mathbf G$. For the
non-flat metrics the $L$-matrices obtained in Maple are
non-symmetric.

If $C_1=a_3-a_2$ and $C_2=a_2$ the eigenvalues $Q_{1,2}$ of this
L-tensor are zeroes of the following function
\[
e(\lambda)={\displaystyle
\frac{\det(L-\lambda)}{\prod_{i=1}^n(\lambda-a_i)}}=\frac{{q_1}}{\lambda-a_1}
+\frac{{q_2}}{\lambda-a_2} +\frac{{1-q_1-q_2}}{\lambda-a_3}
\]
In the redundant coordinates $x_i$ one gets the following definition
of the separation coordinates
\[
\sum_{j=1}^{n+1}\frac{x_j^2}{\lambda-a_j}=
\frac{\prod_{i=1}^n(\lambda-Q_i)}{\prod_{j=1}^{n+1}(\lambda-a_j)}\,,\qquad
\mathrm{with}\qquad \sum_{i=1}^{n+1}x_j^2=1\,.
\]
 These are well-known defining relations for the
spheroconical or elliptic-spherical coordinates. At $n>2$ one gets
the same relations too.

In contrast with the previous examples in this case wehave
non-trivial metric and  $L^i_j(q)\neq L^{ij}(q)$. In order to get
integrals of motion it is more convenient to use recursion relations
(\ref{recc-len-pedr}). The corresponding part of code looks like
\begin{verbatim}
> Nqn:=wcollect(simplify(map2(subs,Ans,Nq))):
> Npn:=wcollect(simplify(map2(subs,Ans,Np))):

> for m from 0 to n do
>  sigma[m]:=coeff(linalg[det](Ln-lambda*ed),lambda,n-m):
> end do:

> unassign(H): H[n+1]:=0: H[0]:=Tn+Vn;
> for m from n-1 by -1 to 1 do
>   Eq:=wcollect(dvalue( d(H[m+1])-
>       subs(Nqn union Npn,d(H[m](var)))-sigma[m+1]*d(H[0]))):
>   Sys:=annul(Eq,[var]):
>   Ans1:=pdsolve(Sys,H[m](var));
>   H[m]:=subs(Ans1,H[m](var)):
> end do:
\end{verbatim}
Of course, we can check the Poisson brackets relations between these
integrals of motion
\begin{verbatim}
> for m from 1 to n-1 do
>   ZERO:=simplify(PB(H[0],H[m]));
> end do;
\end{verbatim}
At $n=2,3$ all the calculations take about one minute and eight
minutes respectively.

\vskip0.3truecm\par\noindent \textbf{Remark:} In contrast with
\cite{bfp03} we directly solve equations (\ref{maple1}-\ref{maple2})
without any additional assumptions about the affine structure of the
solutions. It allows us to prove that there is one unique solution
only.

\subsection{The Jacobi-Calogero inverse-square model}
Configuration space $\mathcal Q$ is  $3$-dimensional Euclidean space
$\mathbb R^3$ with cartesian coordinates $q=(q_1,q_2,q_3)$. The
Hamilton function is given by
\[
H=\sum_{i=1}^3 p_i^2+(q_1-q_2)^{-2}
+(q_2-q_3)^{-2}+(q_3-q_1)^{-2}\,.
\]
For this Hamiltonian  solution of the complete system of equations
\verb"System" depends on four arbitrary parameters
\[
L=\left[\begin{array}{ccc}  q_1^2C_1+2q_1C_2+C_3&
  q_1q_2C_1+(q_1+q_2)C_2+C_4 &   q_1q_3C_1+(q_1+q_3)C_2+C_4 \\[2ex]
  q_1q_2C_1+(q_1+q_2)C_2+C_4 &   q_2^2C_1+
2q_2C_2+C_3 &
  q_2q_3C_1+(q_2+q_3)C_2+C_4 \\[2ex]   q_1q_3C_1+(q_1+q_3)C_2+C_4 &
  q_2q_3C_1+(q_2+q_3)C_2+C_4 &
q_3^2C_1+2q_3C_2+C_3
\end{array}\right]
\]
The eigenvalues of this L-tensor are the separation coordinates
\begin{verbatim}
> Q:=linalg[eigenvalues](L);
\end{verbatim}
The number of free parameters $C_k$ is more  than the dimension of
the Riemannian manifold. It means that  this integrable system is
degenerate or superintegrable system, which admits separation of
variables in some different coordinate systems. Namely, at the
different values of the parameters $C_1,\ldots,C_4$ eigenvalues of
the tensor $\mathbf L$ are oblate spheroidal, prolate spheroidal,
spherical, rotational parabolic and circular cylindrical
coordinates. The detailed discussion may be found in
\cite{rw04,ben00}.

\subsection{An elliptic egg}
Let us consider the Hamiltonian with the rational potential
\[
H=\sum_{i=1}^n p_i^2+{
\frac{c}{\left(1-{\displaystyle\sum_{i=1}^n{\displaystyle\frac{q_i^2}{a_i^2}}}
\right)}}\,,\qquad c,a_i\in\mathbb R\,.
\]
In contrast with the previous examples potential $V(q)$ is a
rational function rather than a polynomial one. This potential has a
singular surface, which is an ellipsoid
\[\sum_{i=1}^n\frac{q_i^2}{a_i^2}=1.\]
For $c >$ 0, this singular surface is repelling inwards, so that all
trajectories starting inside the ellipsoid remain there forever.

At $n>2$ the Maple procedure \verb"pdsolve" can not solve the system
of equations \verb"System" obtained from the both equations
(\ref{maple3}) and (\ref{maple4}) in the reasonable time.

In this case we have to solve equation for the geodesic motion
(\ref{maple3}) and then the equation on  potential (\ref{maple4}).
At $n\leq 10$ it takes just a few minutes.

In our case  solution of the system of equations generated by the
geodesic equation (\ref{maple3}) is equal to
\[
L^i_j(q)=\alpha q_iq_j+ A_{ij}q_i+B_{ij}q_j+C_{ij}\,.
\]
Here coefficients $\alpha, A_{ij}, B_{ij}$ and $C_{ij}$ are
arbitrary constants, $A_{ij}=B_{ji}$ and $C_{ij}=C_{ji}$.

Substituting this solution into the system of equations obtained
from (\ref{maple4}) one gets a system of algebraic equations on the
coefficients $A_{ij}, B_{ij}$ and $C_{ij}$. The standard Maple
program \verb"solve" easy solves this system of algebraic equations
in few seconds. The eigenvalues of the corresponding tensor $\mathbf
L$ are the standard elliptic coordinates (\ref{ell-coord}).

\section{Conclusion}
We present the first part of the symbolic software which builds the
separation coordinates in the Hamilton-Jacobi equation for the
L-systems. The second part of this software is devoted to building
and solving  the corresponding separated equations. We leave this
part to subsequent publications.

Recall, that  Maple file with the first part of the code may be
found in the following URL
\begin{verbatim}
http://www.maplesoft.com/applications/app_center_view.aspx?AID=1686
\end{verbatim}
The latest version of the Maple file may be obtained from authors.

\end{document}